\documentclass[fleqn,useAMS,usenatbib,usegraphicx]{mn2e}
\usepackage{times}
\usepackage{natbib}
\usepackage{epsfig}
\usepackage{rotating}
\usepackage{lscape}
\usepackage{amssymb}
\usepackage{color}

\begin{document}

\def \nhi  {$N_{\rm HI}$}
\def\sci#1{{\; \times \; 10^{#1}}}
\def\cm#1{\, {\rm cm^{#1}}}
\def \mnhi  {N_{\rm HI}}
\def \lya  {Ly$\alpha$}
\def \mlya  {{\rm Ly\alpha}}
\def\hal{H$\alpha$}
\def\smpyr{{\rm M}_{\odot} \, {\rm yr^{-1}}}
\def\msun{{\rm M}_{\odot}}
\def\arcs{\mbox{\ensuremath{^{\prime\prime}}}}
\def\farc{\mbox{\ensuremath{.\!\!\arcs}}}
\def\micron{\hbox{$\mu$m}}
\def\ga{\mathrel{\raise0.35ex\hbox{$\scriptstyle >$}\kern-0.6em
\lower0.40ex\hbox{{$\scriptstyle \sim$}}}}
\def\la{\mathrel{\raise0.35ex\hbox{$\scriptstyle <$}\kern-0.6em
\lower0.40ex\hbox{{$\scriptstyle \sim$}}}}
\def\co{CO~{\it J}=1-0 }
\def\kms{km~s$^{-1}$}
\def\hi{H{\sc i}}
\def\cii{C{\sc ii}-158$\mu$m}
\def\cotwo{CO~{\it J}=2-1 }
\def\cothree{CO~{\it J}=3-2 }
\def\hcm{HCM\,6A}
\def\cosix{CO~{\it J}=6-5 }
\def\hij{high-{\it J}~}
\def\loj{low-{\it J}~}
\def\arcs{\hbox{$^{\prime\prime}$}}
\def\ion#1#2{{#1}\,{\sevensize {#2}}}
\renewcommand{\ion}[2]{\ifmmode 
\mbox{\rm #1\,{\small\uppercase\expandafter{\romannumeral #2}}}
\else\mbox{#1\,{\small\uppercase\expandafter{\romannumeral #2}}}\fi}

\def\etals{{ et al. }\rm}
\def\arcm{\hbox{$^{\prime}$}}

\newcommand{\oneskip}{\vskip \baselineskip}
\newcommand{\annrev}{Annual Review of Astronomy \& Astrophysics}
\newcommand{\araa}{Annual Review of Astronomy \& Astrophysics}
\def\aap{Astronomy \& Astrophysics}
\def\aj{Astronomical Journal}
\def\apj{Astrophysical Journal}
\def\apjl{Astrophysical Journal Letters}
\def\apjs{Astrophysical Journal Supplements}
\def\mnras{Monthly Noticies of the Royal Astronomical Society}
\def\nat{Nature}
\def\pasp{Publications of the Astronomical Society of the Pacific}
\def\prd{Physical Reviews D}
\def\jcap{Journal of Cosmology and Astroparticle Physics}


\title[H$\alpha$ emission in high-$z$ DLAs]{A search for H$\alpha$ emission in high-metallicity damped
  Lyman-$\alpha$ systems at $z \sim 2.4$}

\author[Wang, Kanekar, \& Prochaska]{Wei-Hao Wang$^{1,4}$\thanks{E-mail: whwang@asiaa.sinica.edu.tw},
Nissim~Kanekar$^2\thanks{Ramanujan Fellow}$,
J. Xavier Prochaska$^3$\\
$^1$Academia Sinica Institute of Astronomy and Astrophysics, P.O. Box
23-141, Taipei 10617, Taiwan\\
$^2$National Centre for Radio Astrophysics, Tata Institute of Fundamental Research, Pune 411 007, India\\
$^3$UCO/Lick Observatory, UC Santa Cruz, Santa Cruz, CA 95064, USA\\
$^4$Canada-France-Hawaii Telescope, Kamuela, HA 96743, USA\\
}

\maketitle

\begin{abstract}
We report on a sensitive search for redshifted H$\alpha$ line-emission from three
high-metallicity damped Ly$\alpha$ absorbers (DLAs) at $z \approx 2.4$ 
with the Near-infrared Integral Field Spectrometer (NIFS) on the 
Gemini-North telescope, assisted by the ALTtitude conjugate Adaptive optics 
for the InfraRed (ALTAIR) system with a laser guide star. Within the NIFS 
field-of-view, $\approx 3\farcs22\times2\farcs92$ corresponding to 
$\approx 25$~kpc~$ \times 23$~kpc 
at $z=2.4$, we detect no statistically significant line-emission at the 
expected redshifted H$\alpha$ wavelengths. The measured root-mean-square noise 
fluctuations in $0\farcs4$ apertures are $1-3\times10^{-18}$~erg~s$^{-1}$~cm$^{-2}$.
Our analysis of simulated, compact, line-emitting sources yields stringent 
limits on the 
star-formation rates (SFRs) of the three DLAs, $< 2.2$~M$_{\odot}$~yr$^{-1}$ ($3\sigma$) 
for two absorbers, and $< 11$~M$_{\odot}$~yr$^{-1}$ ($3\sigma$) for the third, at all impact 
parameters within $\approx 12.5$~kpc to the quasar sightline at the DLA redshift. For the 
third absorber, the SFR limit is $< 4.4$~M$_\odot$~yr$^{-1}$ for locations away from the 
quasar sightline. These 
results demonstrate the potential of adaptive optics-assisted, integral 
field unit searches for galaxies associated with high-$z$ DLAs. 
\end{abstract}


\section{Introduction}

The damped \lya\ systems \citep[DLAs;][]{wolfe86,wolfe05} define the class 
of absorption-line systems discovered in the rest-frame UV spectra of distant
quasars, with \ion{H}{1} column densities $\mnhi \ge 2\sci{20} \cm{-2}$ as measured
from the analysis of damping wings in the \lya\ profile. These large \nhi\ values, 
characteristic of gas in the interstellar medium of modern, star-forming galaxies
\citep[e.g.][]{walter08}, have led researchers to associate high-$z$ DLAs with young 
galaxies in the early Universe.  This association is well supported by theoretical 
models of galaxy formation \citep[e.g.][]{haehnelt98,maller01,bird14} and recent 
numerical simulations that predict large \ion{H}{1} surface densities extending out 
to tens of kpc from active, star-forming regions \citep[e.g.][]{rahmati13,fumagalli14a} 
and even to distances of $\approx 200$~kpc from the centres of the most massive halos 
\citep{pontzen08}. Further empirical evidence connecting high-$z$ DLAs to galaxies 
includes the omnipresence of heavy elements in the gas typically at levels exceeding 
estimates for the intergalactic medium which indicates previous, if not recent, 
pollution by stars \citep[e.g.][]{pettini94,pettini97,prochaska03a,penprase10,rafelski12}.
At low redshifts, where one may more sensitively search for associated galaxies, 
several studies have identified galaxy counterparts to the absorbers 
\citep[e.g.][]{lebrun97,rao03,chen05}.

Detailed analysis of DLA absorption lines offer unparalleled insight into the 
interstellar medium (ISM) of these galaxies, including their chemical abundance 
patterns, gas temperature, molecular content, dust depletion, and kinematics 
\citep[e.g.][]{pettini97,prochaska97,kanekar03,ledoux03,dessauges06,noterdaeme08,kanekar14}.
Indeed, DLA studies provided the first and (still) the most comprehensive view of the 
ISM at high redshifts.  To properly place this wealth of ISM measurements within
galaxy formation theories, however, it is critical to connect the gas to the 
stars and to the processes of star formation. Establishing this connection directly 
by observing the stars and/or \ion{H}{2} regions, however, has proven to be 
remarkably challenging.

Early works searched first for \lya\ emission, both in the core of the \ion{H}{1} 
\lya\ absorption profile and with narrow-band imaging or, more recently, with integral 
field spectroscopy
\citep[e.g.][]{smith89,hunstead90,wolfe92,moller93,kulkarni06,christensen07}.
These provided a few rare detections \citep[e.g.][]{djorkowski96,fynbo99,moller04} and mostly 
upper limits to the star-formation rates (SFRs), $< \sim 10\; \smpyr$, albeit 
subject to substantial systematic uncertainties related to dust depletion and \lya\ 
radiative transfer. The community then searched sensitively for far-UV continuum emission in 
broad-band imaging, applying the Lyman-break technique when possible to pre-select 
candidates at high redshifts \citep[e.g.][]{moller93,warren01,prochaska02}; these surveys 
were largely unsuccessful, with only a few positive detections. This implied inherently 
faint ultraviolet emission and/or that the absorbing galaxies were drowned out by the glare 
of the background quasar. More recently, \cite{fumagalli10} introduced a novel technique 
to address the problem of detecting a faint DLA close to a bright quasar, using a 
high-$\mnhi$ absorption system at a redshift close to that of the background quasar 
to ``block'' its far-ultraviolet light, thus allowing a sensitive search for emission from 
a second, lower-redshift DLA along the sightline. Unfortunately, even this technique has 
so far yielded null results, albeit yielding more stringent constraints, 
SFR~$\le 0.1 - 0.3\; \smpyr$ at all impact parameters \citep{fumagalli14b,fumagalli15}.

\begin{table*}
\caption{DLA Sample \label{tab:DLAs}}
\begin{tabular}{lcccccccc}
\hline
QSO & RA (J2000) & DEC (J2000) & $z_{\rm em}$ & $z_{\rm DLA}$ & \nhi &
[Z/H]$^\ast$ & References \\
\hline
QSO0201+365  & 02:04:55.60 & +36:49:18 & 2.912 & 2.4628  & 20.38 &
$-0.29$ 	& 1, 2 \\
QSO0311+430  & 03:14:43.60 & +43:14:05.1 & 2.870 & 2.2898 & 20.30 & 
$-0.49$ & 3, 4, 5 \\
QSO2343+125 & 23:46:28.22 & +12:48:59.9 & 2.763	& 2.4313  & 20.34 &
$-0.54^\dagger$  & 6, 7, 8 \\
\hline
\end{tabular}
\vskip 0.05in
Notes to the table:
$^\ast$~The metallicity estimates are based on [Zn/H] (QSO0201+365) 
and [Si/H]~(QSO0311+430 and QSO2343+125).\\
$^\dagger$~The metallicity of the $z = 2.4313$ DLA towards QSO2343+125
is from the HIRES spectrum of \citet{lu98}. We have re-analysed these
data and obtain a slightly higher, albeit consistent, metallicity of 
[Si/H]$= -0.47$.  We note that \citet{ledoux06} obtained
a lower metallicity from the Zn{\sc ii} lines, [Zn/H]$= -0.92 \pm 0.07$,
while \citet{dessauges04} used S{\sc ii} lines to obtain 
[S/H]$=-0.7 \pm 0.1$ (consistent with the metallicity estimate 
of \citet{lu98}). \\
References: (1)~\citealp{prochaska96}; (2)~\citealp{prochaska01b};
(3)~\citealp{york07}; (4)~\citealp{ellison08}; (5)~\citealp{kanekar14};
(6)~\citealp{lu98}; (7)~\citealp{dessauges04}; (8)~\citealp{prochaska07}.

\end{table*}

Somewhat stymied, the community has made two changes in the observational strategy:
(1)~focus the effort on the most heavily enriched DLAs under the expectation 
that these will be associated with brighter galaxies having higher SFRs and/or 
larger stellar masses \citep[e.g.][]{moller04,ledoux06}, and (2)~search for the
 strong nebular emission lines associated with \ion{H}{2} regions (e.g.\ \ion{O}{3}, 
\hal). Earlier studies using the latter approach yielded relatively weak
upper limits on the SFRs of DLAs, $\lesssim 10 - 100\, \smpyr$ at $z \gtrsim 2$
\citep[e.g.][when converted to the LCDM cosmology used in this paper]{bechtold98,mannucci98,bunker99,kulkarni00}. 
However, this method
has recently been significantly advanced by technical progress, i.e.\ the 
commissioning of higher throughput near-infrared (near-IR) spectrometers on 
large-aperture telescopes.  Several of these include integral field units (IFUs) 
which permit a ``blind'' search for emission in a modest field around each DLA. These 
new efforts, mostly with the Very Large Telescope (VLT), have met with somewhat greater 
success: $\approx 10$ counterparts to DLAs at $z \gtrsim 2$ have recently been discovered, 
at impact parameters from the quasar sightlines in the range
$\approx 1- 23$\,kpc\footnote{Throughout this paper, we will use a flat 
$\Lambda$-cold dark matter cosmology, with $\Omega_m = 0.315$, $\Omega_\Lambda = 
0.685$ and H$_0 = 67.3$~\kms~Mpc$^{-1}$ \citep{planck13}.}, and with 
SFRs~$10-20 \; \smpyr$ \citep[e.g.][]{fynbo10,fynbo11,fynbo13,peroux11,peroux12,noterdaeme12}. 
There have also been a few recent detections of \lya\ emission in the trough of 
the DLA \lya\ absorption line
\citep[e.g.][]{moller04,hennawi09,kulkarni12,krogager12,noterdaeme14}.

One is scientifically motivated to search for star-formation from DLAs for 
several reasons beyond mere studies of this absorber population. In contrast 
to the multitude of surveys of star-forming galaxies at $z \sim 2$ 
\citep[e.g.][]{daddi04,erb06,forster-schreiber09}, one may link the gas
directly to the stars and star-forming regions in these young
galaxies.  For \ion{H}{1} in particular, absorption-line analysis
represents the only means for assessing the surface density of gas
around these galaxies and thereby an estimate of the gaseous extent.

Studies of the star-forming regions of DLAs also offer a complementary assessment 
of emerging trends in $z \sim 2$ galaxies between stellar mass $M^*$, metallicity 
and star-formation rate. Large spectroscopic surveys have revealed that 
star-forming galaxies follow a relatively tight main sequence, in which the SFR 
correlates with the stellar mass \citep[e.g.][]{noeske07,daddi07,rodighiero11}. 
This has inspired a series of simple models of galaxy formation within
the $\Lambda$CDM framework that consider the balance of gas accretion,
star formation, and the outflow of material by feedback processes 
\citep[e.g.][]{dekel09,lilly13,forbes14}.  Observers have also explored
second-order relations in the galaxy properties, emphasizing the existence 
of a fundamental metallicity relation (FMR) where the SFR is systematically 
higher in lower metallicity galaxies at a fixed mass
\citep{mannucci10,bothwell13,stott13}.   
With the DLAs, one may examine whether absorption-selected galaxies
follow the same trends as these traditional samples.  In addition,
one may further test the simple models by incorporating the ISM 
properties constrained directly by DLA observations (e.g.\ \ion{H}{1}
extent, ISM metallicity, star formation history, gas temperature, etc).

In this manuscript, we report on our own project to search for \hal\ emission from high-$z$ 
DLAs with the Near-infrared Integral Field Spectrometer \citep[NIFS; ][]{mcgregor03}
on the Gemini-North telescope.   We pre-selected a set of DLAs with high metallicity and with
absorption redshifts well-suited to the detection of nebular emission lines (e.g.\ 
strong sky lines were avoided). In addition, we used the ALTtitude conjugate Adaptive optics 
for the InfraRed (ALTAIR) with a laser guide star to isolate the light of the background 
quasar and maximize the signal from these (presumably) small galaxies. We note 
red that \citet{jorgenson14} also employed a similar approach, with 
laser-guided adaptive optics and an IFU with the Keck\,I telescope to analyze 
a previously known galaxy counterpart to a $z \approx 2.3543$~DLA with [\ion{O}{3}] 
and \hal\ emission.

This paper is organized as follows: $\S$~\ref{sec:obs} describes the DLA sample, the 
observational programme, and the details of the data reduction.  The primary results are 
given in $\S$~\ref{sec:results} and we conclude with a discussion of our non-detections, 
similar searches in the literature, and future efforts in $\S$~\ref{sec:discuss}.

\section{Observations and Data Reduction}
\label{sec:obs}

From the set of known  DLAs at $z \sim 2$ in 2007, with accurately measured
metallicities \citep[e.g.][]{ledoux06,prochaska07}, we considered all systems with 
enrichment exceeding $\approx 1/3$ solar abundance.  We further restricted this 
set to have good visibility from Mauna Kea and chose five systems with absorption 
redshifts $z_{\rm abs}$ that placed their \hal\ lines at favorable wavelengths for
deep, near-IR spectroscopy (i.e.\ avoiding bright sky lines and strong telluric 
absorption). Unfortunately, two of the DLAs were observed with incorrect wavelength 
settings, so we were left with three absorbers; Table~\ref{tab:DLAs} summarizes their 
properties.

We used NIFS on Gemini-North to search for redshifted \hal\ emission from the three DLAs 
in 2008, with ALTAIR and laser guide stars to improve the point spread function (PSF).
Nearby stars or the QSO itself were used for the tip-tilt mode correction in the laser 
guiding. The data were acquired in queue-observing mode during semester~2008B under 
Program~ID GN-2008B-Q-61. The observing conditions were photometric with
natural seeing better than $0\farcs65$ in the optical $V$-band as measured by the ALTAIR
wave-front sensor.

We used the $K$ grating to search for \hal\ emission from the three DLAs at $z \approx 2.4$.
For each target, the observing band covered a wavelength range of approximately 4000~\AA, 
centered at the expected DLA \hal\ wavelength (see Table~\ref{tab1}).  The NIFS field-of-view
is approximately $3\arcsec$ on a side, corresponding to $\approx 25$\,kpc at $z=2.4$.
Each exposure was set to 560~seconds, a compromise between minimizing the effects of 
instrumental flexure and read-out noise. The exposures were dithered along R.A.\ and Dec.\ 
with $\pm0\farcs15$ offsets to obtain better estimates of the sky background.  
This offset was chosen to match the expected PSF of $\sim0\farcs15$. However, the actually 
achieved PSF is worse than this (see below), leading to a less optimal sky subtraction for slightly
extended objects.
We did not 
take off-target exposures to independently measure the sky background. A0V stars at similar 
airmass were observed immediately before and/or after the target observations with an identical 
instrumental setup to estimate telluric absorption and to perform flux calibration. 
Images of Ar and Xe arc lamps were obtained to provide wavelength calibration.

\begin{table*}
\centering
\caption{Observing Logs\label{tab1}}
\begin{tabular}{lcccccccc}
\hline
Field & {Date (UT)} & {$\lambda_{obs}$ ($\mu$m)} & {Telluric Stars} & {$T_{\rm exp}$ (sec)} & {Mag$_{\rm tt}$} & {$d_{\rm tt}$} \\
\hline
QSO2343+125 	& 2008 Oct 4  & 2.2519 & HD 21501 		& 4480 & $V=17.0$   & $0\arcsec$ \\ 
QSO0201+365  	& 2008 Oct 7  & 2.2726 & HD 1561, HD 21038 	& 5600 & $R = 16.2$ & $19\farcs0$ \\
QSO0311+430 	& 2008 Oct 7  & 2.1590 & HD 12559 		& 2800 & $F=16.0$   &  $26\farcs4$ \\ 
\hline
\end{tabular}
\begin{center}
Mag$_{\rm tt}$ and $d_{\rm tt}$ are the magnitudes of the tip-tilt guide stars and their
angular distances to the targets, respectively.\\
\end{center}
\end{table*}

\begin{figure}
\centering
\includegraphics[width=0.45\textwidth]{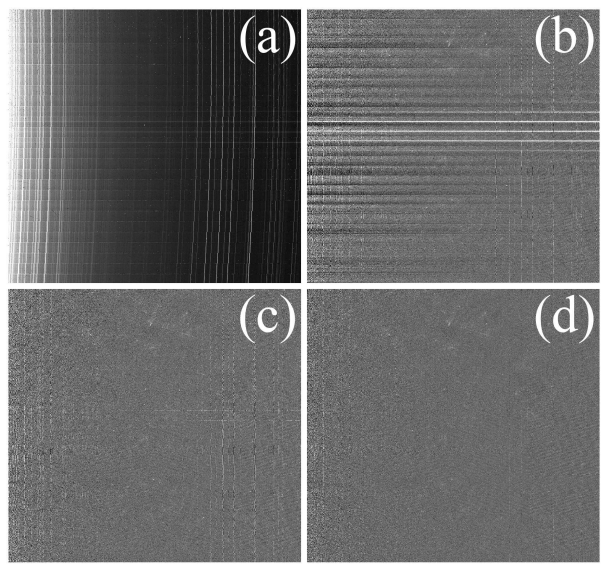}
\caption{Example of our NIFS data processing (see main text for details). In all panels, 
the horizontal and vertical directions are the spectral and spatial directions, respectively.
(a)~shows an image after dark subtraction and flat fielding.  
(b)~shows the result after the vertical sky lines are horizontally aligned and 
subtracted. The bright horizontal lines near the middle of the image are the QSO 
continuum, and the broader horizontal bands are residual sky background. 
(c)~shows the result after the subtraction of a smooth function, column by column and 
then row by row. 
(d)~shows the result after cosmic ray removal and subtraction of an image that is 
median-combined from all dithered exposures.}
\label{fig:example}
\end{figure}

The data were reduced with our own programmes written in the Interactive Data Language 
environment. The images were first dark subtracted and flattened; Figure~\ref{fig:example}a 
shows an example of one processed frame. As is evident from the figure, the NIFS lenslets 
project spectra onto the detector with different wavelength zeropoints. To begin sky 
subtraction, we first registered the sky lines by applying a rigid pixel offset to each row
derived from a 1-D cross-correlation, after a $2\times$ re-sampling in the spectral dimension.  
Because the $2\times$ re-sampled spectral pixels have a size of 1.03~\AA, this rigid pixel
offset does not introduce a wavelength error larger than $0.5$~\AA.  

\begin{figure}
\centering
\includegraphics[width=0.45\textwidth]{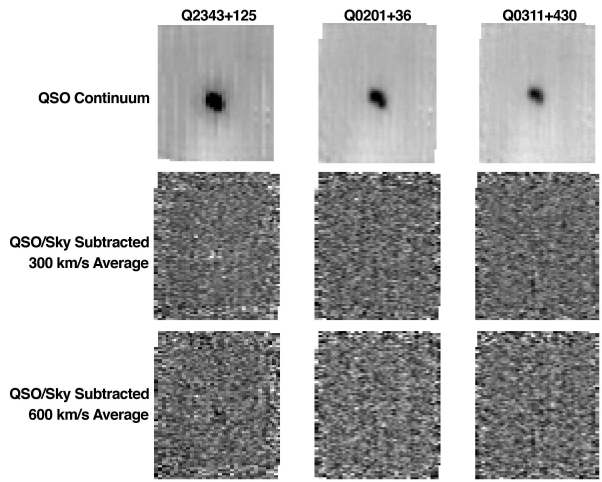}
\caption{Results of our Gemini-NIFS search for \hal\ emission from three metal-rich DLAs.  
Each panel has an angular size of $3\farcs22\times2\farcs92$ (corresponding to 
$\sim25 {\rm kpc} \times 23$~kpc at $z=2.4$) and has an inverted gray scale. 
The top row shows continuum images of the three background QSOs. The middle and bottom 
rows show averaged images within 10 and 22~spectral pixels (corresponding to 
$\sim 300$~km~s$^{-1}$ and $\sim 600$ km s$^{-1}$), respectively, centered at the 
expected redshifted \hal\ wavelengths, after subtraction of the QSOs and the sky 
background. No evidence of \hal\ emission is found in these images or in the 
data cubes.
\label{fig:images}}
\end{figure}

After the rows of spectra were aligned, a constant sky background was subtracted column 
by column (Figure~\ref{fig:example}b). At this stage, this simple sky-subtraction was sufficient 
to allow detection of the QSOs over the entire spectral range, in each 560s exposure.  
The original NIFS 2-D images have a spatial pixel size of $0\farcs103\times0\farcs04$ and 
spectral pixel sizes of $2.07$~\AA. The images were reformatted and re-sampled into 3-D cubes of 
$0\farcs052\times0\farcs04\times2.07$~\AA\ pixel sizes. The cubes were collapsed along the 
spectral axes to form continuum images of the QSOs (Figure~\ref{fig:images}, top panels). 
Centroids of the quasar emission were used 
to derive dither offsets for the multiple exposures with a precision that is much smaller 
than the pixel size. We also estimated the PSF provided by the ALTAIR adaptive optics (AO) system 
to lie in the range $0.2-0.3''$ (full width at half maximum; FWHM).  

To sensitively search for the DLA H$\alpha$ emission, one requires a much higher fidelity 
approach to sky subtraction and cosmic ray removal.  Figure~\ref{fig:example}b shows obvious 
residuals at the brightest sky lines and also ``ripples'' in the spatial dimension that we 
associate with spatial variations in the sky background and/or non-uniformities in the 
lenslets that have not been captured by the flat-fielding procedure. We also wished to search 
sensitively for \hal\ emission close to, or even at, the position of each QSO. Therefore, we 
fitted a 5th or 6th-order polynomial, column by column and then row-by-row, to each NIFS 2D image 
and subtracted the fitted values. Figure~\ref{fig:example}c shows a resultant image. This 
procedure is designed to remove slowly varying flux fluctuations related to the background and
the QSO without eliminating narrow features such as emission lines and cosmic rays. For the 
latter, we flagged all pixels with fluxes $8\sigma$ above the background and verified (by eye) 
that none of these could be consistent with an emission feature. Lastly, we median-combined 
the five to ten dithered images without alignment to generate a final residual map of the sky 
background including the effects of ``hot'' pixels.  This residual map was subtracted from 
each image (e.g.\ Figure~\ref{fig:example}d) and these images were reformatted to 3-D cubes. 
The dithered cubes of each target were then aligned and combined with sigma-clipping to form 
the final data cube for each field.

The spectral images of the Ar/Xe arc lamps and the A0V stars were processed separately.
Sky-subtraction was not carried out here since the sources are much brighter than the 
structure in the sky background. The extracted Ar/Xe spectra were used to derive 
wavelength solutions, accurate to $\sim1$~\AA\ over the observed spectral ranges, 
which is sufficient for our science. The extracted A0V spectra were used to derive 
telluric absorptions and perform flux calibration. First, we masked the 2.166 $\mu$m 
hydrogen absorption lines on the A0V spectra and interpolated over the masked regions in 
the continua. The spectra were then compared with the model spectrum of Vega 
\citep{kurucz79} to derive the telluric absorptions as a function of wavelength. 
Absorption corrections of $\approx 1\%$ were applied to both the data cubes of the 
targets and the A0V spectra. Finally, the absolute flux calibrations were estimated 
with the 2MASS $K_S$ fluxes of the A0V stars and their observed spectral fluxes in 
the 2MASS $K_S$ passband ($\sim2.17$ $\mu$m). The flux zero points derived from the 
four A0V stars under three different spectral settings agree very well with each 
other (to within $\sim3\%$), suggesting that our flux calibration is accurate to 
within 5\%.

\section{Results}
\label{sec:results}

\begin{figure}
\centering
\includegraphics[width=0.45\textwidth]{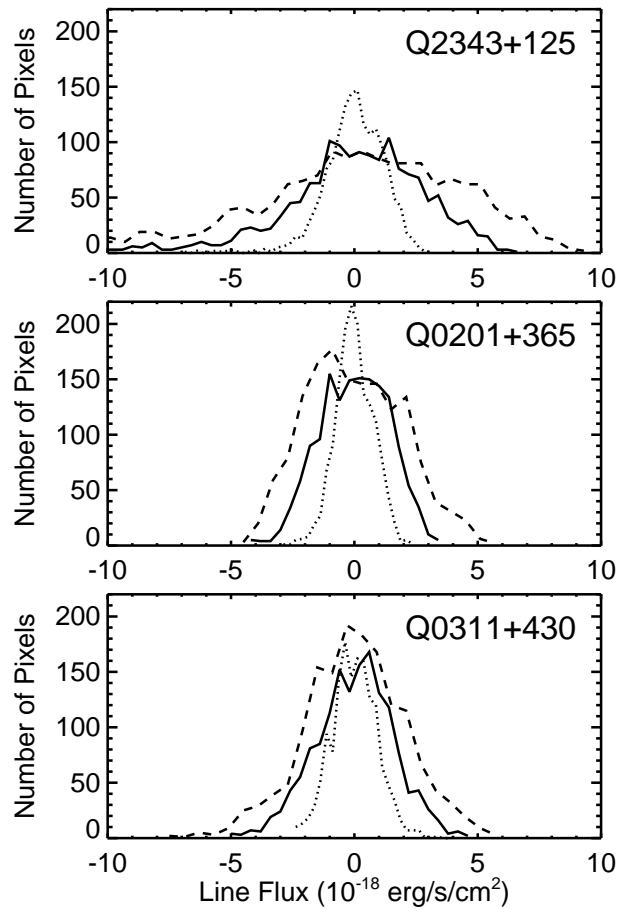}
\caption{Histograms of the fluxes measured in $0\farcs2$ (dotted), $0\farcs4$ (solid), and 
$0\farcs6$ (dashed) square apertures across  the images in each field. The measurements were
made within a $\sim300$ km s$^{-1}$ velocity range, centered on the expected redshifted \hal\ wavelength.
}
\label{fig:hist}
\end{figure}

The final reduced images are shown in Figure~\ref{fig:images}.  Each panel is 
$3\farcs22\times2\farcs92$ in size, corresponding to approximately
$25$~kpc~$\times \, 23$~kpc
 at $z=2.4$.  The top row shows the stacked QSO continuum images from an intermediate 
stage before processing (see $\S$~\ref{sec:obs}).  Here, the sky subtraction is very simple 
and residual sky structures are apparent. In the case of QSO0201+365 and QSO0311+430, the 
tip-tilt guide stars are quite far away from the QSO and we only obtained partially 
corrected PSFs from ALTAIR.  The measured PSF FWHMs are $0\farcs24\times0\farcs17$ for 
QSO2343+125, $0\farcs36\times0\farcs23$ for QSO0201+365, and $0\farcs31\times0\farcs23$ 
for QSO0311+430.  These are all much better than the natural seeing but worse than what 
ALTAIR can achieve in the best cases ($\sim0\farcs1$).  The asymmetric PSFs are likely to 
arise from the laser-guided AO performance, as the self-guided observations of the 
A0V stars do not show such asymmetry.

The middle row of the figure shows stacked images spanning 10 spectral pixels 
(corresponding to $\sim 300$~km~s$^{-1}$) centered at the expected redshifted 
\hal\ wavelength for each DLA, after further subtraction of the QSO and sky background 
(see $\S$~\ref{sec:obs}).  Both the QSO and the sky were cleanly subtracted across the 
images, enabling the search for faint \hal\ emission even along the sightline to each 
QSO. There is no evidence of significant H$\alpha$ emission in any of the three DLAs. 
We also carefully searched the 3-D image cubes over a broader wavelength range, in 
images before and after cosmic ray removal, and in individual unstacked exposures. 
We found no evidence for emission in any of these cases.   To search for very broad 
emission, we also created images averaged over 22 spectral pixels (corresponding to 
$\sim 600$ km s$^{-1}$, bottom row of Figure~\ref{fig:images}); no emission was 
detected at this coarser resolution.  For the discussion below, we will assume 
a line width of $\sim 300$~km~s$^{-1}$, i.e. we will use the images of the middle 
row in Fig.~\ref{fig:images}.

During the data reduction process, we median-combined the dithered exposures to 
form residual images for better sky subtraction, using dither offsets between 
exposures of $\approx 0\farcs15$. This means that the DLA \hal\ emission might be 
partially subtracted if it is smooth and extended over angular scales of 
$\gtrsim \farcs3$. We did not see any hint of such extended emission in the images 
before the median sky subtraction, nor in the median-combined sky image.  Furthermore, 
such spatially-extended, but spectrally-narrow, emission (extended over 
$> 2.5$~kpc at $z=2.4$) appears unlikely \citep[e.g.][]{fynbo11}.  We conclude that 
the three DLAs of our sample do not produce strong \hal\ emission within 
$\sim 12.5$~kpc of the QSO sightline.

We used the non-detections of \hal\ emission to estimate upper limits to the 
\hal\ flux from each DLA.  We measured the background flux fluctuations in the 
300~km~s$^{-1}$ images shown in the middle row of Figure~\ref{fig:images} in 
$0\farcs4$ square apertures across each image. The $0\farcs4$ size is larger than 
the PSF FWHM in all three images, and corresponds to 3.3~kpc at $z=2.4$. 
Figure~\ref{fig:hist} shows a histogram of these fluxes, as well as results 
for other aperture sizes. The measured root-mean-square (RMS) noise fluctuations 
in $0\farcs4$ apertures are $2.86\times10^{-18}$~erg~s$^{-1}$~cm$^{-2}$, 
$1.38\times10^{-18}$~erg~s$^{-1}$~cm$^{-2}$, and 
$1.57\times10^{-18}$ erg s$^{-1}$ cm$^{-2}$, in the images for QSO2343+125, 
QSO0201+365, and QSO0311+430, respectively (see Figure~\ref{fig:hist}). The 
above values can be converted to limits on the SFR via the expression
$\mathrm{SFR} = 7.9 \times 10^{-42} L_{\rm H\alpha}$~M$_{\odot}$~yr$^{-1}$~(erg~s$^{-1}$)$^{-1}$
\citep{kennicutt98}, where a Salpeter initial mass function is assumed.  
The results are  $1\sigma$ SFR limits of $1.11$~M$_\odot$~yr$^{-1}$, 
$0.55$~M$_\odot$~yr$^{-1}$, and $0.53$~M$_\odot$~yr$^{-1}$, for QSO2343+125, 
QSO0201+365, and QSO0311+430, respectively.  For a line width of 600~km~s$^{-1}$,
the RMS noise and the SFR limits increase approximately by a factor of $\sqrt{2}$.

Our relatively small dither offsets ($0\farcs15$) and the non-ideal PSFs lead to 20\%--30\% 
(depending on the source size and PSF) flux losses in the median sky subtraction. As a 
result, the sensitivity to \hal\ line emission is somewhat worse than the typically quoted 
$3\sigma$ upper limits based on the above $1\sigma$ error estimates. In order to better 
understand the systematics arising from our observing and data analysis procedure, and to 
obtain reliable constraints on the SFRs, we created artificial line emitting sources in 
each of the raw images of our target sources and processed these images identically to the 
actual images.  We assumed intrinsically extended emitters with various sizes of Gaussians
of up to $0\farcs2$ FWHM.  We then convolved the source images with the PSF derived from 
the QSO images. The sources were conservatively assumed to have flat velocity profiles 
over $10-15$~\AA. ($\approx 100-200$~km~sec$^{-1}$) and were randomly placed in the raw 
images with a range of fluxes.  We consider a simulated source as recovered if
it passes our inspection and its flux is detected at $\geq 3\sigma$ significance. We found 
that we confidently recover the simulated sources in our reduction when the input flux 
is $\approx 4$ times larger than the formal RMS noise. This  is consistent with 
the above-mentioned 20\%--30\% flux loss caused by the broad PSF.  We thus regard this as
a $3\sigma$ confidence detection limit. Note that this is true even when the simulated sources 
have very small impact parameters to the QSO sightline, except for the case 
of QSO2343+125. QSO2343+125 has broad emission lines in our observed spectral cube, making 
our quasar subtraction slightly poorer.  In this case, we can only detect the simulated 
line source along the QSO sightline when it has $\gtrsim 10\sigma$ significance. We 
therefore conclude that our observations constrain the star formation rates to 
SFR~$< 2.2$~M$_{\odot}$~yr$^{-1}$ at all impact parameters for the DLAs toward QSO0201+365 
and QSO0311+430, and SFR~$<4.4$~M$_{\odot}$~yr$^{-1}$ at separations $>0\farcs5$ for the 
DLA towards QSO2343+125, with a poorer limit of SFR~$< 11$~M$_{\odot}$~yr$^{-1}$ close 
to the QSO for the last system. We note that all of these limits ignore dust obscuration 
beyond the modest correction associated with the empirical relation of \citet{kennicutt98}.
Our observations  would thus underestimate the SFR limits in the case of modest 
amount of extinction, and would be insensitive to dusty star-forming regions, for which the 
\hal\ line emission might be highly obscured.

\section{Discussion}
\label{sec:discuss}

The primary aim of searches for line emission from high-$z$ DLAs is the direct 
identification of their galaxy counterparts, so as to estimate the typical size, 
mass and SFR of absorption-selected galaxies. Non-detections of \hal\ emission only 
allow us to place limits on the SFR, assuming that the emission region of the DLA 
host lies within the search area. In addition, however, there has been much recent 
interest in whether high-$z$ absorption-selected galaxies also have 
a mass-metallicity relation similar to that seen in high-$z$ emission-selected galaxies 
\citep[e.g.][]{tremonti04,erb06}. The existence of such a relation in DLAs is suggested by 
the correlations found between DLA metallicity and both the velocity width of low-ionization 
metal lines \citep[][]{wolfe98,ledoux06,neeleman13} and the rest equivalent width of the 
\ion{Si}{2}$\lambda$1526 line \citep{prochaska08}. Similarly, the anti-correlation detected 
between DLA metallicity and gas spin temperature \citep{kanekar09c,ellison12,kanekar14} 
also supports such a mass-metallicity relation, as higher mass DLAs are expected to 
have larger fractions of cold gas and hence, lower spin temperatures \citep{kanekar01a}.
To this end, even non-detections of \hal\ emission in high-metallicity DLAs are of much 
interest, as these may argue against the existence of a mass-metallicity relation 
in DLAs given the star-forming main sequence of $z \sim 2$ galaxies
\citep[e.g.][]{daddi07}.

Before we consider the repercussions of our results, we briefly discuss other 
recent searches for line emission from high-$z$ DLAs. There have been two broad 
strategies that have been successful in either detecting, or placing strong 
constraints on, \lya, \hal, \ion{O}{2} or \ion{O}{3} line emission. (1)~\citet{fynbo10} 
pioneered an approach based on using VLT X-shooter slit spectroscopy, covering a 
large wavelength range ($3100$~\AA~$-2.5\mu$m) to simultaneously search for 
a number of lines, and using 3~slit position angles centred on the QSO location 
\citep[see also][]{fynbo11,krogager12,fynbo13}. (2)~Conversely, \citet{peroux11} 
used IFU spectroscopy with the Spectrograph for Integral Field Observations in 
the Near-Infrared (SINFONI) onboard the VLT to search for redshifted \hal\ emission,
similar to our approach \citep[see also][]{bouche12,peroux12}. Fynbo et al. also 
targeted high-metallicity DLAs at $z \gtrsim 2$, while Peroux et al. and Bouche et al. 
targeted DLAs, sub-DLAs and strong \ion{Mg}{2} absorbers of all metallicities, 
centred at two redshifts, $z \approx 1$ and $z \approx 2$. 

There are pros and cons to the two literature approaches, as well as to our own 
approach. The X-shooter-based strategy of \citet{fynbo10} has the advantage 
that a number of lines are simultaneously targeted, implying fewer problems
due to night sky lines and sky subtraction. Also, the use of 3 slit angles reduces 
the likelihood that the DLA might be missed by the slit. However, the likelihood of 
missing the DLA increases with distance from the QSO location (as the slits are 
centred on the QSO), implying a bias towards absorbers at low impact parameter, 
$\lesssim 1''$. 

The VLT-SINFONI IFU searches for \hal\ emission by \citet{peroux12} and \citet{bouche12} 
have the advantage of being sensitive to \hal\ emission from a relatively large area (field 
of view $\approx 10\times 10$~arcsec$^2$) around the QSO sightline. However, AO was used 
(with nearby guide stars) for only a few DLAs and even these only yielded a seeing 
$\geq 0\farcs5$.  The relatively poor seeing and the degradation in the signal-to-noise
ratio at the QSO location (due to the higher background) implies that the sensitivity 
of such searches is lower close to the QSO. Non-detections of \hal\ emission might 
hence arise at low impact parameters, $\lesssim 0\farcs5$ \citep[although see][]{bouche12}. 

In the case of our own search, the use of AO implies excellent seeing and high 
sensitivity to \hal\ emission even at the QSO location for two of the targets, 
and reasonable sensitivity for the third target. However, the relatively small 
NIFS field of view implies that we are only sensitive to emission for impact
parameters $\lesssim 12.5$~kpc. Thus, \hal\ emission from large disks (of radius 
$\gg 10$~kpc), where the \lya\ absorption occurs in the outskirts of the galaxy 
while the \hal\ emission is dominated by the central region might be missed in 
our search. Note that all three DLAs of our sample have relatively low \hi\ column 
densities, $< 3 \times 10^{20} \cm{-2}$, as might be expected in the outer regions of 
a galaxy. 

Moving to detections, and focusing on DLAs at $z \gtrsim 2$ (i.e. excluding low-$z$ 
DLAs and \ion{Mg}{2} absorbers without \nhi\ estimates), the X-shooter method has 
led to the detection of line emission from five high-metallicity DLAs at $z > 2$ 
\citep{fynbo10,krogager12,fynbo13}. Non-detections from this approach have so far not 
been reported. Perhaps unsurprisingly, four of the five detections \citep[see][]{krogager12} 
have been at relatively low impact parameters, $\lesssim 0\farcs8$.

In the case of VLT-SINFONI IFU spectroscopy, \citet{peroux12} report only a single 
detection of \hal\ emission in 10 DLAs and 2 sub-DLAs at 
$z \gtrsim 2$. Further, even the sole DLA of their sample with a detection of 
\hal\ emission (at $z = 2.3543$ towards J2222$-$0946) had earlier been detected 
in \hal, \lya\, \ion{O}{2} and \ion{O}{3} lines in X-shooter spectroscopy \citep{fynbo10}.
This system also has an intermediate impact parameter, $b \approx 0\farcs8$, larger than 
the seeing during the VLT-SINFONI observations \citep[$\approx 0\farcs6$; ][]{peroux12}.
Note that only two DLAs of the \citet{peroux12} sample are ``high-metallicity'' systems, 
with [Z/H]~$\gtrsim -0.5$; one of these is the absorber towards J2222$-$0946, 
while the other, at $z = 2.2100$ towards QSO2059$-$0528, yielded an \hal\ non-detection,
with an upper limit to the SFR, $< 3 \smpyr$.\footnote{Note that different authors 
have used different spatial and spectral resolutions, as well as significances, when 
quoting upper limits. Here, we adjusted their values to a fiducial $3\sigma$ 
significance and have converted to the same H$\alpha$ calibration adopted here.
Differences between the adopted cosmologies have negligible impact on the results.}

\citet{bouche12} also used VLT-SINFONI to search for \hal\ emission in a set of absorbers 
at $z \gtrsim 2$, selected either due to strong \ion{Mg}{2} absorption or due to identification 
as DLAs or sub-DLAs. We will here only consider the subset of 10 absorbers, 5 DLAs and 5 sub-DLAs,
with \nhi\ estimates. These yielded two detections of \hal\ emission, one system again being
the ubiquitous $z = 2.3543$ DLA towards J2222$-$0946. The sole other detection was in the $z = 2.3288$ 
DLA towards HE2243$-$6031, an intermediate metallicity absorber ([Zn/H]~$= -1.10 \pm 0.05$; 
\citealp{lopez02}), detected at a large impact parameter, $\approx 26.5$~kpc, with an SFR of $17 \smpyr$ 
\citep[see also][]{bouche13}. The seven non-detections yielded typical SFR limits of 
$3.2 \; \smpyr$.

Finally, our own search in three DLAs yielded no \hal\ detections, with strong 
SFR constraints of $2.2-4.4 \smpyr$ within a distance of 12.5~kpc to the QSO sightline 
(excepting the QSO location in one field). All three DLAs have high metallicities, 
[Z/H]~$\geq -0.54$, within the top 10\% of DLAs at all redshifts \citep[e.g.][]{rafelski12}. 

It thus appears that, despite a number of studies with 8m-class telescopes, there have been 
very few IFU detections of \hal\ emission in DLAs at $z \gtrsim 2$. Specifically, combining our 
results with those of \citet{peroux12} and \citet{bouche12}, there has been only a single ``new'' 
detection of \hal\ emission, and a corresponding identification of a galaxy counterpart, and 21 
non-detections from IFU searches. Conversely, there have been five identifications of galaxy 
counterparts via the X-shooter slit spectroscopy approach \citep{fynbo10,fynbo11,krogager12}. 
The target DLA samples of the two types of searches have been very different, with the IFU 
searches mostly targeting low-metallicity DLAs (18/22) and the X-shooter spectroscopy focusing 
on high-metallicity DLAs. This suggests that high-metallicity DLAs have higher 
SFRs, as would be expected from a mass-metallicity relation in DLAs
\citep[e.g.][]{moller04,ledoux06,moller13}, together with the 
observed SFR-$M^*$ main sequence. 

On the other hand, the three high-metallicity DLAs of our sample were
not detected in a sensitive IFU search. These have metallicities
similar to those of three of the X-shooter targets in
\citet{krogager12}, and would have been detected in our search for
similar SFRs ($\gtrsim 2.2 - 4.4 \smpyr$) and impact parameters ($\lesssim
8$ kpc).   Adopting the SFR-$M^*$ relation of \cite{daddi07} and
extrapolating it to lower SFRs, our limits imply $M^* < 10^{9.1}
\msun$.   Is such a low mass inconsistent with the DLA metallicities? 
Evaluating the FMR of \citet{mannucci10}, which is based on emission lines,
with $M^* = 10^{9.1} \msun$
and SFR~$= 2.2 \smpyr$, we recover 12+log(O/H)~$\approx 8.3$\,dex.
This value is roughly consistent with the DLA absorption metallicity, especially 
given the systematic uncertainty of nebular-line metallicity measurements
inherent to the FMR and the possibility of abundance gradients.

Of course, the number of high-metallicity DLAs that have been searched for \hal\ 
or nebular emission is still quite small and it is possible that observational 
biases might have caused some of the non-detections. For example, in the model of 
\citet{fynbo08}, high-metallicity DLAs are expected to arise in large, luminous 
galaxies, with statistically larger impact parameters. The \hal\ non-detections 
in our searches might then have been due to the limited NIFS field of view. We 
conclude that, although the paucity of detections in IFU searches is certainly 
curious, all three approaches (VLT-X-shooter, Gemini-NIFS and VLT-SINFONI) have 
biases (see above) that would affect the fraction of detections of \hal\ or 
nebular emission, and hence the inferred SFR limits. We suggest that the present 
approaches should be augmented by (1)~combining wide-field IFU searches with the 
use of AO with laser guide stars, so that \hal\ emission cannot be hidden under 
the QSO emission or missed by being outside the field of view, and (2)~following 
up X-shooter non-detections with more slit angles, to rule out the possibility 
that non-detections arise because the nebular emission is not covered by the slit. 
Finally, given the biases in the current searches, we also feel that
it is premature to interpret the present SFR data to either support or
rule out a mass-metallicity relation in damped Ly$\alpha$ systems.

\section*{Acknowledgements}

This work is based on observations obtained at the Gemini Observatory, which is 
operated by the Association of 
Universities for Research in Astronomy, Inc., under a cooperative agreement with 
the NSF on behalf of the Gemini partnership: the National Science Foundation (United 
States), the Science and Technology Facilities Council (United Kingdom), the National 
Research Council (Canada), CONICYT (Chile), the Australian Research Council 
(Australia), Minist\'{e}rio da Ci\^{e}ncia, Tecnologia e Inova\c{c}\~{a}o (Brazil) 
and Ministerio de Ciencia, Tecnolog\'{i}a e Innovaci\'{o}n Productiva (Argentina).
WHW is partially supported by the Ministry of Science and Technology of Taiwan grant
102-2119-M-001-007-MY3. NK acknowledges support from the Department of Science and 
Technology through a Ramanujan Fellowship. JXP is partly supported by NSF grant AST-1109447.
The authors would like to dedicate this paper to the memory of Arthur M. Wolfe, 
who pioneered studies of damped Ly$\alpha$ systems for more than three decades.

\bibliographystyle{mn2e}

\begin{thebibliography}{}

\bibitem[\protect\citeauthoryear{{Bechtold}, {Elston}, {Yee}, {Ellingson} \&
  {Cutri}}{{Bechtold} et~al.}{1998}]{bechtold98}
{Bechtold} J.,  {Elston} R.,  {Yee} H.~K.~C.,  {Ellingson} E.,    {Cutri}
  R.~M.,  1998, in {D'Odorico} S.,  {Fontana} A.,   {Giallongo} E.,  eds, The
  Young Universe: Galaxy Formation and Evolution at Intermediate and High
  Redshift Vol.~146 of Astronomical Society of the Pacific Conference Series,
  {Star Formation in High Redshift Galaxies}.
p.~241

\bibitem[\protect\citeauthoryear{{Bird}, {Vogelsberger}, {Haehnelt}, {Sijacki},
  {Genel}, {Torrey}, {Springel} \& {Hernquist}}{{Bird} et~al.}{2014}]{bird14}
{Bird} S.,  {Vogelsberger} M.,  {Haehnelt} M.,  {Sijacki} D.,  {Genel} S.,
  {Torrey} P.,  {Springel} V.,    {Hernquist} L.,  2014, MNRAS, 445, 2313

\bibitem[\protect\citeauthoryear{{Bothwell}, {Maiolino}, {Kennicutt}, {Cresci},
  {Mannucci}, {Marconi} \& {Cicone}}{{Bothwell} et~al.}{2013}]{bothwell13}
{Bothwell} M.~S.,  {Maiolino} R.,  {Kennicutt} R.,  {Cresci} G.,  {Mannucci}
  F.,  {Marconi} A.,    {Cicone} C.,  2013, MNRAS, 433, 1425

\bibitem[\protect\citeauthoryear{{Bouch{\'e}}, {Murphy}, {Kacprzak},
  {P{\'e}roux}, {Contini}, {Martin} \& {Dessauges-Zavadsky}}{{Bouch{\'e}}
  et~al.}{2013}]{bouche13}
{Bouch{\'e}} N.,  {Murphy} M.~T.,  {Kacprzak} G.~G.,  {P{\'e}roux} C.,
  {Contini} T.,  {Martin} C.~L.,    {Dessauges-Zavadsky} M.,  2013, Science,
  341, 50

\bibitem[\protect\citeauthoryear{{Bouch{\'e}}, {Murphy}, {P{\'e}roux},
  {Contini}, {Martin}, {F{\"o}rster Schreiber}, {Genzel}, {Lutz}, {Gillessen},
  {Tacconi}, {Davies} \& {Eisenhauer}}{{Bouch{\'e}} et~al.}{2012}]{bouche12}
{Bouch{\'e}} N. et al., 2012, MNRAS, 419, 2

\bibitem[\protect\citeauthoryear{{Bunker}, {Warren}, {Clements}, {Williger} \&
  {Hewett}}{{Bunker} et~al.}{1999}]{bunker99}
{Bunker} A.~J.,  {Warren} S.~J.,  {Clements} D.~L.,  {Williger} G.~M.,
  {Hewett} P.~C.,  1999, MNRAS, 309, 875

\bibitem[\protect\citeauthoryear{{Chen}, {Kennicutt} Jr. \& {Rauch}}{{Chen}
  et~al.}{2005}]{chen05}
{Chen} H.-W.,  {Kennicutt} Jr. R.~C.,    {Rauch} M.,  2005, ApJ, 620, 703

\bibitem[\protect\citeauthoryear{{Christensen}, {Wisotzki}, {Roth},
  {S{\'a}nchez}, {Kelz} \& {Jahnke}}{{Christensen}
  et~al.}{2007}]{christensen07}
{Christensen} L.,  {Wisotzki} L.,  {Roth} M.~M.,  {S{\'a}nchez} S.~F.,  {Kelz}
  A.,    {Jahnke} K.,  2007, A\&A, 468, 587

\bibitem[\protect\citeauthoryear{{Daddi}, {Cimatti}, {Renzini}, {Fontana},
  {Mignoli}, {Pozzetti}, {Tozzi} \& {Zamorani}}{{Daddi} et~al.}{2004}]{daddi04}
{Daddi} E.,  {Cimatti} A.,  {Renzini} A.,  {Fontana} A.,  {Mignoli} M.,
  {Pozzetti} L.,  {Tozzi} P.,    {Zamorani} G.,  2004, ApJ, 617, 746

\bibitem[\protect\citeauthoryear{{Daddi}, {Dickinson}, {Morrison}, {Chary},
  {Cimatti}, {Elbaz}, {Frayer}, {Renzini}, {Pope}, {Alexander}, {Bauer},
  {Giavalisco}, {Huynh}, {Kurk} \& {Mignoli}}{{Daddi} et~al.}{2007}]{daddi07}
{Daddi} E. et al., 2007, ApJ, 670, 156

\bibitem[\protect\citeauthoryear{{Dekel}, {Birnboim}, {Engel}, {Freundlich},
  {Goerdt}, {Mumcuoglu}, {Neistein}, {Pichon}, {Teyssier} \& {Zinger}}{{Dekel}
  et~al.}{2009}]{dekel09}
{Dekel} A.  et al., 2009, Nature, 457, 451

\bibitem[\protect\citeauthoryear{{Dessauges-Zavadsky}, {Calura}, {Prochaska},
  {D'Odorico} \& {Matteucci}}{{Dessauges-Zavadsky} et~al.}{2004}]{dessauges04}
{Dessauges-Zavadsky} M.,  {Calura} F.,  {Prochaska} J.~X.,  {D'Odorico} S.,
  {Matteucci} F.,  2004, A\&A, 416, 79

\bibitem[\protect\citeauthoryear{{Dessauges-Zavadsky}, {Prochaska},
  {D'Odorico}, {Calura} \& {Matteucci}}{{Dessauges-Zavadsky}
  et~al.}{2006}]{dessauges06}
{Dessauges-Zavadsky} M.,  {Prochaska} J.~X.,  {D'Odorico} S.,  {Calura} F.,
  {Matteucci} F.,  2006, A\&A, 445, 93

\bibitem[\protect\citeauthoryear{{Djorgovski}, {Pahre}, {Bechtold} \&
  {Elston}}{{Djorgovski} et~al.}{1996}]{djorkowski96}
{Djorgovski} S.~G.,  {Pahre} M.~A.,  {Bechtold} J.,    {Elston} R.,  1996,
  Nature, 382, 234

\bibitem[\protect\citeauthoryear{{Ellison}, {Kanekar}, {Prochaska}, {Momjian}
  \& {Worseck}}{{Ellison} et~al.}{2012}]{ellison12}
{Ellison} S.~L.,  {Kanekar} N.,  {Prochaska} J.~X.,  {Momjian} E.,    {Worseck}
  G.,  2012, MNRAS, 424, 293

\bibitem[\protect\citeauthoryear{{Ellison}, {York}, {Pettini} \&
  {Kanekar}}{{Ellison} et~al.}{2008}]{ellison08}
{Ellison} S.~L.,  {York} B.~A.,  {Pettini} M.,    {Kanekar} N.,  2008, MNRAS,
  388, 1349

\bibitem[\protect\citeauthoryear{{Erb}, {Shapley}, {Pettini}, {Steidel},
  {Reddy} \& {Adelberger}}{{Erb} et~al.}{2006}]{erb06}
{Erb} D.~K.,  {Shapley} A.~E.,  {Pettini} M.,  {Steidel} C.~C.,  {Reddy} N.~A.,
     {Adelberger} K.~L.,  2006, ApJ, 644, 813

\bibitem[\protect\citeauthoryear{{Forbes}, {Krumholz}, {Burkert} \&
  {Dekel}}{{Forbes} et~al.}{2014}]{forbes14}
{Forbes} J.~C.,  {Krumholz} M.~R.,  {Burkert} A.,    {Dekel} A.,  2014, MNRAS,
  438, 1552

\bibitem[\protect\citeauthoryear{{F{\"o}rster Schreiber} {et
  al.}}{{F{\"o}rster Schreiber} {et al.}}{2009}]{forster-schreiber09}
{F{\"o}rster Schreiber} N.~M. et al., 2009, ApJ, 706, 1364

\bibitem[\protect\citeauthoryear{{Fumagalli}, {Hennawi}, {Prochaska}, {Kasen},
  {Dekel}, {Ceverino} \& {Primack}}{{Fumagalli} et~al.}{2014a}]{fumagalli14a}
{Fumagalli} M.,  {Hennawi} J.~F.,  {Prochaska} J.~X.,  {Kasen} D.,  {Dekel} A.,
   {Ceverino} D.,    {Primack} J.,  2014a, ApJ, 780, 74

\bibitem[\protect\citeauthoryear{{Fumagalli}, {O'Meara}, {Prochaska}, {Kanekar}
  \& {Wolfe}}{{Fumagalli} et~al.}{2014b}]{fumagalli14b}
{Fumagalli} M.,  {O'Meara} J.~M.,  {Prochaska} J.~X.,  {Kanekar} N.,  {Wolfe} A.~M.,
    2014b, MNRAS, 444, 1282

\bibitem[\protect\citeauthoryear{{Fumagalli}, {O'Meara}, {Prochaska}, {Rafelski} \& 
 {Kanekar}}{{Fumagalli} et~al.}{2015}]{fumagalli15}
{Fumagalli} M.,  {O'Meara} J.~M.,  {Prochaska} J.~X., {Rafelski} M., {Kanekar} N., 
    2015, 446, 3178


\bibitem[\protect\citeauthoryear{{Fumagalli}, {O'Meara}, {Prochaska} \&
  {Kanekar}}{{Fumagalli} et~al.}{2010}]{fumagalli10}
{Fumagalli} M.,  {O'Meara} J.~M.,  {Prochaska} J.~X.,    {Kanekar} N.,  2010,
  MNRAS, 408, 362

\bibitem[\protect\citeauthoryear{{Fynbo}, {M{\o}ller} \& {Warren}}{{Fynbo}
  et~al.}{1999}]{fynbo99}
{Fynbo} J.~P.,  {M{\o}ller} P.,    {Warren} S.~J.,  1999, MNRAS, 305, 849

\bibitem[\protect\citeauthoryear{{Fynbo}, {Geier}, {Christensen}, {Gallazzi},
  {Krogager}, {Kr{\"u}hler}, {Ledoux}, {Maund}, {M{\o}ller}, {Noterdaeme},
  {Rivera-Thorsen} \& {Vestergaard}}{{Fynbo} et~al.}{2013}]{fynbo13}
{Fynbo} J.~P.~U. et al., 2013, MNRAS, 436, 361

\bibitem[\protect\citeauthoryear{{Fynbo}, {Laursen}, {Ledoux}, {M{\o}ller},
  {Durgapal}, {Goldoni}, {Gullberg}, {Kaper}, {Maund}, {Noterdaeme},
  {{\"O}stlin}, {Strandet}, {Toft}, {Vreeswijk} \& {Zafar}}{{Fynbo}
  et~al.}{2010}]{fynbo10}
{Fynbo} J.~P.~U.  et al., 2010, MNRAS, 408, 2128

\bibitem[\protect\citeauthoryear{{Fynbo}, {Ledoux}, {Noterdaeme},
  {Christensen}, {M{\o}ller}, {Durgapal}, {Goldoni}, {Kaper}, {Krogager},
  {Laursen}, {Maund}, {Milvang-Jensen}, {Okoshi}, {Rasmussen}, {Thorsen},
  {Toft} \& {Zafar}}{{Fynbo} et~al.}{2011}]{fynbo11}
{Fynbo} J.~P.~U.  et al., 2011, MNRAS, 413, 2481

\bibitem[\protect\citeauthoryear{{Fynbo}, {Prochaska}, {Sommer-Larsen},
  {Dessauges-Zavadsky} \& {M{\o}ller}}{{Fynbo} et~al.}{2008}]{fynbo08}
{Fynbo} J.~P.~U.,  {Prochaska} J.~X.,  {Sommer-Larsen} J.,
  {Dessauges-Zavadsky} M.,    {M{\o}ller} P.,  2008, ApJ, 683, 321

\bibitem[\protect\citeauthoryear{Haehnelt, Steinmetz \& Rauch}{Haehnelt
  et~al.}{1998}]{haehnelt98}
Haehnelt M.~G.,  Steinmetz M.,    Rauch M.,  1998, ApJ, 495, 64

\bibitem[\protect\citeauthoryear{{Hennawi}, {Prochaska}, {Kollmeier} \&
  {Zheng}}{{Hennawi} et~al.}{2009}]{hennawi09}
{Hennawi} J.~F.,  {Prochaska} J.~X.,  {Kollmeier} J.,    {Zheng} Z.,  2009,
  ApJL, 693, L49

\bibitem[\protect\citeauthoryear{{Hunstead}, {Pettini} \&
  {Fletcher}}{{Hunstead} et~al.}{1990}]{hunstead90}
{Hunstead} R.~W.,  {Pettini} M.,    {Fletcher} A.~B.,  1990, ApJ, 356, 23

\bibitem[\protect\citeauthoryear{{Jorgenson} \& {Wolfe}}{{Jorgenson} \&
  {Wolfe}}{2014}]{jorgenson14}
{Jorgenson} R.~A.,  {Wolfe} A.~M.,  2014, ApJ, 785, 16

\bibitem[\protect\citeauthoryear{{Kanekar} \& {Chengalur}}{{Kanekar} \&
  {Chengalur}}{2001}]{kanekar01a}
{Kanekar} N.,  {Chengalur} J.~N.,  2001, A\&A, 369, 42

\bibitem[\protect\citeauthoryear{{Kanekar} \& {Chengalur}}{{Kanekar} \&
  {Chengalur}}{2003}]{kanekar03}
{Kanekar} N.,  {Chengalur} J.~N.,  2003, A\&A, 399, 857

\bibitem[\protect\citeauthoryear{{Kanekar}, {Prochaska}, {Smette}, {Ellison},
  {Ryan-Weber}, {Momjian}, {Briggs}, {Lane}, {Chengalur}, {Delafosse}, {Grave},
  {Jacobsen} \& {de Bruyn}}{{Kanekar} et~al.}{2014}]{kanekar14}
{Kanekar} N.  et al., 2014, MNRAS, 438, 2131

\bibitem[\protect\citeauthoryear{{Kanekar}, {Smette}, {Briggs} \&
  {Chengalur}}{{Kanekar} et~al.}{2009}]{kanekar09c}
{Kanekar} N.,  {Smette} A.,  {Briggs} F.~H.,    {Chengalur} J.~N.,  2009, ApJ,
  705, L40

\bibitem[\protect\citeauthoryear{{Kennicutt}
  Jr.}{{Kennicutt}}{1998}]{kennicutt98}
{Kennicutt} Jr. R.~C.,  1998, ApJ, 498, 541

\bibitem[\protect\citeauthoryear{{Krogager}, {Fynbo}, {M{\o}ller}, {Ledoux},
  {Noterdaeme}, {Christensen}, {Milvang-Jensen} \& {Sparre}}{{Krogager}
  et~al.}{2012}]{krogager12}
{Krogager} J.-K.,  {Fynbo} J.~P.~U.,  {M{\o}ller} P.,  {Ledoux} C.,
  {Noterdaeme} P.,  {Christensen} L.,  {Milvang-Jensen} B.,    {Sparre} M.,
  2012, MNRAS, 424, L1

\bibitem[\protect\citeauthoryear{{Kulkarni}, {Hill}, {Schneider}, {Weymann},
  {Storrie-Lombardi}, {Rieke}, {Thompson} \& {Jannuzi}}{{Kulkarni}
  et~al.}{2000}]{kulkarni00}
{Kulkarni} V.~P.,  {Hill} J.~M.,  {Schneider} G.,  {Weymann} R.~J.,
  {Storrie-Lombardi} L.~J.,  {Rieke} M.~J.,  {Thompson} R.~I.,    {Jannuzi}
  B.~T.,  2000, ApJ, 536, 36

\bibitem[\protect\citeauthoryear{{Kulkarni}, {Meiring}, {Som}, {P{\'e}roux},
  {York}, {Khare} \& {Lauroesch}}{{Kulkarni} et~al.}{2012}]{kulkarni12}
{Kulkarni} V.~P.,  {Meiring} J.,  {Som} D.,  {P{\'e}roux} C.,  {York} D.~G.,
  {Khare} P.,    {Lauroesch} J.~T.,  2012, ApJ, 749, 176

\bibitem[\protect\citeauthoryear{{Kulkarni}, {Woodgate}, {York}, {Thatte},
  {Meiring}, {Palunas} \& {Wassell}}{{Kulkarni} et~al.}{2006}]{kulkarni06}
{Kulkarni} V.~P.,  {Woodgate} B.~E.,  {York} D.~G.,  {Thatte} D.~G.,  {Meiring}
  J.,  {Palunas} P.,    {Wassell} E.,  2006, ApJ, 636, 30

\bibitem[\protect\citeauthoryear{{Kurucz}}{{Kurucz}}{1979}]{kurucz79}
{Kurucz} R.~L.,  1979, ApJS, 40, 1

\bibitem[\protect\citeauthoryear{{le Brun}, Bergeron, Boiss\'{e} \&
  Deharveng}{{le Brun} et~al.}{1997}]{lebrun97}
{le Brun} V.,  Bergeron J.,  Boiss\'{e} P.,    Deharveng J.-M.,  1997, A\&A,
  321, 733

\bibitem[\protect\citeauthoryear{{Ledoux}, {Petitjean}, {Fynbo}, {M{\o}ller} \&
  {Srianand}}{{Ledoux} et~al.}{2006}]{ledoux06}
{Ledoux} C.,  {Petitjean} P.,  {Fynbo} J.~P.~U.,  {M{\o}ller} P.,    {Srianand}
  R.,  2006, A\&A, 457, 71

\bibitem[\protect\citeauthoryear{{Ledoux}, {Petitjean} \& {Srianand}}{{Ledoux}
  et~al.}{2003}]{ledoux03}
{Ledoux} C.,  {Petitjean} P.,    {Srianand} R.,  2003, MNRAS, 346, 209

\bibitem[\protect\citeauthoryear{{Lilly}, {Carollo}, {Pipino}, {Renzini} \&
  {Peng}}{{Lilly} et~al.}{2013}]{lilly13}
{Lilly} S.~J.,  {Carollo} C.~M.,  {Pipino} A.,  {Renzini} A.,    {Peng} Y.,
  2013, ApJ, 772, 119

\bibitem[\protect\citeauthoryear{{Lopez}, {Reimers}, {D'Odorico} \&
  {Prochaska}}{{Lopez} et~al.}{2002}]{lopez02}
{Lopez} S.,  {Reimers} D.,  {D'Odorico} S.,    {Prochaska} J.~X.,  2002, A\&A,
  385, 778

\bibitem[\protect\citeauthoryear{{Lu}, {Sargent} \& {Barlow}}{{Lu}
  et~al.}{1998}]{lu98}
{Lu} L.,  {Sargent} W.~L.~W.,    {Barlow} T.~A.,  1998, AJ, 115, 55

\bibitem[\protect\citeauthoryear{{Maller}, {Prochaska}, {Somerville} \&
  {Primack}}{{Maller} et~al.}{2001}]{maller01}
{Maller} A.~H.,  {Prochaska} J.~X.,  {Somerville} R.~S.,    {Primack} J.~R.,
  2001, MNRAS, 326, 1475

\bibitem[\protect\citeauthoryear{{Mannucci}, {Cresci}, {Maiolino}, {Marconi} \&
  {Gnerucci}}{{Mannucci} et~al.}{2010}]{mannucci10}
{Mannucci} F.,  {Cresci} G.,  {Maiolino} R.,  {Marconi} A.,    {Gnerucci} A.,
  2010, MNRAS, 408, 2115

\bibitem[\protect\citeauthoryear{{Mannucci}, {Thompson}, {Beckwith} \&
  {Williger}}{{Mannucci} et~al.}{1998}]{mannucci98}
{Mannucci} F.,  {Thompson} D.,  {Beckwith} S.~V.~W.,    {Williger} G.~M.,
  1998, ApJL, 501, L11

\bibitem[\protect\citeauthoryear{{McGregor}, {Hart}, {Conroy}, {Pfitzner},
  {Bloxham}, {Jones}, {Downing}, {Dawson}, {Young}, {Jarnyk} \& {Van
  Harmelen}}{{McGregor} et~al.}{2003}]{mcgregor03}
{McGregor} P.~J. et al., 2003, in {Iye} M.,  {Moorwood} A.~F.~M.,  eds,
  Instrument Design and Performance for Optical/Infrared Ground-based
  Telescopes Vol.~4841 of Society of Photo-Optical Instrumentation Engineers
  (SPIE) Conference Series, {Gemini near-infrared integral field spectrograph
  (NIFS)}.
p.~1581

\bibitem[\protect\citeauthoryear{{M{\o}ller}, {Fynbo} \& {Fall}}{{M{\o}ller}
  et~al.}{2004}]{moller04}
{M{\o}ller} P.,  {Fynbo} J.~P.~U.,    {Fall} S.~M.,  2004, A\&A, 422, L33

\bibitem[\protect\citeauthoryear{{M{\o}ller}, {Fynbo}, {Ledoux} \&
  {Nilsson}}{{M{\o}ller} et~al.}{2013}]{moller13}
{M{\o}ller} P.,  {Fynbo} J.~P.~U.,  {Ledoux} C.,    {Nilsson} K.~K.,  2013,
  MNRAS, 430, 2680

\bibitem[\protect\citeauthoryear{{M{\o}ller} \& {Warren}}{{M{\o}ller} \&
  {Warren}}{1993}]{moller93}
{M{\o}ller} P.,  {Warren} S.~J.,  1993, A\&A, 270, 43

\bibitem[\protect\citeauthoryear{{Neeleman}, {Wolfe}, {Prochaska} \&
  {Rafelski}}{{Neeleman} et~al.}{2013}]{neeleman13}
{Neeleman} M.,  {Wolfe} A.~M.,  {Prochaska} J.~X.,    {Rafelski} M.,  2013,
  ApJ, 769, 54

\bibitem[\protect\citeauthoryear{{Noeske} {et al.}}{{Noeske} {et
  al.}}{2007}]{noeske07}
{Noeske} K.~G. {et al.}, 2007, ApJ, 660, L43

\bibitem[\protect\citeauthoryear{{Noterdaeme}, {Laursen}, {Petitjean},
  {Vergani}, {Maureira}, {Ledoux}, {Fynbo}, {L{\'o}pez} \&
  {Srianand}}{{Noterdaeme} et~al.}{2012}]{noterdaeme12}
{Noterdaeme} P. et al., 2012, A\&A, 540, 63

\bibitem[\protect\citeauthoryear{{Noterdaeme}, {Ledoux}, {Petitjean} \&
  {Srianand}}{{Noterdaeme} et~al.}{2008}]{noterdaeme08}
{Noterdaeme} P.,  {Ledoux} C.,  {Petitjean} P.,    {Srianand} R.,  2008, A\&A,
  481, 327

\bibitem[\protect\citeauthoryear{{Noterdaeme}, {Petitjean}, {P{\^a}ris}, {Cai},
  {Finley}, {Ge}, {Pieri} \& {York}}{{Noterdaeme} et~al.}{2014}]{noterdaeme14}
{Noterdaeme} P.,  {Petitjean} P.,  {P{\^a}ris} I.,  {Cai} Z.,  {Finley} H.,
  {Ge} J.,  {Pieri} M.~M.,    {York} D.~G.,  2014, A\&A, 566, A24

\bibitem[\protect\citeauthoryear{{Penprase}, {Prochaska}, {Sargent},
  {Toro-Martinez} \& {Beeler}}{{Penprase} et~al.}{2010}]{penprase10}
{Penprase} B.~E.,  {Prochaska} J.~X.,  {Sargent} W.~L.~W.,  {Toro-Martinez} I.,
     {Beeler} D.~J.,  2010, ApJ, 721, 1

\bibitem[\protect\citeauthoryear{{P{\'e}roux}, {Bouch{\'e}}, {Kulkarni}, {York}
  \& {Vladilo}}{{P{\'e}roux} et~al.}{2011}]{peroux11}
{P{\'e}roux} C.,  {Bouch{\'e}} N.,  {Kulkarni} V.~P.,  {York} D.~G.,
  {Vladilo} G.,  2011, MNRAS, 410, 2237

\bibitem[\protect\citeauthoryear{{P{\'e}roux}, {Bouch{\'e}}, {Kulkarni}, {York}
  \& {Vladilo}}{{P{\'e}roux} et~al.}{2012}]{peroux12}
{P{\'e}roux} C.,  {Bouch{\'e}} N.,  {Kulkarni} V.~P.,  {York} D.~G.,
  {Vladilo} G.,  2012, MNRAS, 419, 3060

\bibitem[\protect\citeauthoryear{{Pettini}, {Smith}, {Hunstead} \&
  {King}}{{Pettini} et~al.}{1994}]{pettini94}
{Pettini} M.,  {Smith} L.~J.,  {Hunstead} R.~W.,    {King} D.~L.,  1994, ApJ,
  426, 79

\bibitem[\protect\citeauthoryear{{Pettini}, {Smith}, {King} \&
  {Hunstead}}{{Pettini} et~al.}{1997}]{pettini97}
{Pettini} M.,  {Smith} L.~J.,  {King} D.~L.,    {Hunstead} R.~W.,  1997, ApJ,
  486, 665

\bibitem[\protect\citeauthoryear{{Planck Collaboration}}{{Planck Collaboration}
{}}{2013}]{planck13} {Planck Collaboration} 2013, arxiv:1303.5076

\bibitem[\protect\citeauthoryear{{Pontzen}, {Governato}, {Pettini}, {Booth}, {Stinson},
   {Wadsley}, {Brooks}, {Quinn} \& {Haehnelt}}{{Pontzen} et~al.}{2008}]{pontzen08}
{Pontzen} A. et al., 2008, MNRAS, 390, 1349

\bibitem[\protect\citeauthoryear{{Prochaska}, {Chen}, {Wolfe},
  {Dessauges-Zavadsky} \& {Bloom}}{{Prochaska} et~al.}{2008}]{prochaska08}
{Prochaska} J.~X.,  {Chen} H.-W.,  {Wolfe} A.~M.,  {Dessauges-Zavadsky} M.,
  {Bloom} J.~S.,  2008, ApJ, 672, 59

\bibitem[\protect\citeauthoryear{{Prochaska}, {Gawiser}, {Wolfe}, {Castro} \&
  {Djorgovski}}{{Prochaska} et~al.}{2003}]{prochaska03a}
{Prochaska} J.~X.,  {Gawiser} E.,  {Wolfe} A.~M.,  {Castro} S.,    {Djorgovski}
  S.~G.,  2003, ApJ, 595, L9

\bibitem[\protect\citeauthoryear{{Prochaska}, {Gawiser}, {Wolfe},
  {Quirrenbach}, {Lanzetta}, {Chen}, {Cooke} \& {Yahata}}{{Prochaska}
  et~al.}{2002}]{prochaska02}
{Prochaska} J.~X.,  {Gawiser} E.,  {Wolfe} A.~M.,  {Quirrenbach} A.,
  {Lanzetta} K.~M.,  {Chen} H.-W.,  {Cooke} J.,    {Yahata} N.,  2002, AJ, 123,
  2206

\bibitem[\protect\citeauthoryear{{Prochaska} \& {Wolfe}}{{Prochaska} \&
  {Wolfe}}{1996}]{prochaska96}
{Prochaska} J.~X.,  {Wolfe} A.~M.,  1996, ApJ, 470, 403

\bibitem[\protect\citeauthoryear{Prochaska \& Wolfe}{Prochaska \&
  Wolfe}{1997}]{prochaska97}
Prochaska J.~X.,  Wolfe A.~M.,  1997, ApJ, 487, 73

\bibitem[\protect\citeauthoryear{Prochaska, Wolfe, Howk, Gawiser, Burles \&
  Cooke}{Prochaska et~al.}{2007}]{prochaska07}
Prochaska J.~X.,  Wolfe A.~M.,  Howk J.~C.,  Gawiser E.,  Burles S.~M.,
  Cooke J.,  2007, ApJS, 171, 29

\bibitem[\protect\citeauthoryear{{Prochaska}, {Wolfe}, {Tytler}, {Burles},
  {Cooke}, {Gawiser}, {Kirkman}, {O'Meara} \& {Storrie-Lombardi}}{{Prochaska}
  et~al.}{2001}]{prochaska01b}
{Prochaska} J.~X. et al., 2001, ApJS, 137, 21

\bibitem[\protect\citeauthoryear{{Rafelski}, {Wolfe}, {Prochaska}, {Neeleman}
  \& {Mendez}}{{Rafelski} et~al.}{2012}]{rafelski12}
{Rafelski} M.,  {Wolfe} A.~M.,  {Prochaska} J.~X.,  {Neeleman} M.,    {Mendez}
  A.~J.,  2012, ApJ, 755, 89

\bibitem[\protect\citeauthoryear{{Rahmati}, {Pawlik}, {Rai{\v c}evic} \&
  {Schaye}}{{Rahmati} et~al.}{2013}]{rahmati13}
{Rahmati} A.,  {Pawlik} A.~H.,  {Rai{\v c}evic} M.,    {Schaye} J.,  2013,
  MNRAS, 430, 2427

\bibitem[\protect\citeauthoryear{{Rao}, {Nestor}, {Turnshek}, {Lane}, {Monier}
  \& {Bergeron}}{{Rao} et~al.}{2003}]{rao03}
{Rao} S.~M.,  {Nestor} D.~B.,  {Turnshek} D.~A.,  {Lane} W.~M.,  {Monier}
  E.~M.,    {Bergeron} J.,  2003, ApJ, 595, 94

\bibitem[\protect\citeauthoryear{{Rodighiero} {et al.}}{{Rodighiero} {et
  al.}}{2011}]{rodighiero11}
{Rodighiero} G. et al., 2011, ApJ, 739, L40

\bibitem[\protect\citeauthoryear{{Smith}, {Cohen}, {Burns}, {Moore} \&
  {Uchida}}{{Smith} et~al.}{1989}]{smith89}
{Smith} H.~E.,  {Cohen} R.~D.,  {Burns} J.~E.,  {Moore} D.~J.,    {Uchida}
  B.~A.,  1989, ApJ, 347, 87

\bibitem[\protect\citeauthoryear{{Stott}, {Sobral}, {Bower}, {Smail}, {Best},
  {Matsuda}, {Hayashi}, {Geach} \& {Kodama}}{{Stott} et~al.}{2013}]{stott13}
{Stott} J.~P.,  {Sobral} D.,  {Bower} R.,  {Smail} I.,  {Best} P.~N.,
  {Matsuda} Y.,  {Hayashi} M.,  {Geach} J.~E.,    {Kodama} T.,  2013, MNRAS,
  436, 1130

\bibitem[\protect\citeauthoryear{{Tremonti}, {Heckman}, {Kauffmann},
  {Brinchmann}, {Charlot}, {White}, {Seibert}, {Peng}, {Schlegel}, {Uomoto},
  {Fukugita} \& {Brinkmann}}{{Tremonti} et~al.}{2004}]{tremonti04}
{Tremonti} C.~A. et al., 2004, ApJ, 613, 898

\bibitem[\protect\citeauthoryear{{Walter}, {Brinks}, {de Blok}, {Bigiel},
  {Kennicutt}, {Jr.}, {Thornley} \& {Leroy}}{{Walter} et~al.}{2008}]{walter08}
{Walter} F.,  {Brinks} E.,  {de Blok} W.~J.~G.,  {Bigiel} F.,  {Kennicutt}
  R.~C.,  {Jr.} {Thornley} M.~D.,    {Leroy} A.~K.,  2008, AJ, 136, 2563

\bibitem[\protect\citeauthoryear{{Warren}, {M{\o}ller}, {Fall} \&
  {Jakobsen}}{{Warren} et~al.}{2001}]{warren01}
{Warren} S.~J.,  {M{\o}ller} P.,  {Fall} S.~M.,    {Jakobsen} P.,  2001, MNRAS,
  326, 759

\bibitem[\protect\citeauthoryear{Wolfe, Gawiser \& Prochaska}{Wolfe
  et~al.}{2005}]{wolfe05}
Wolfe A.~M.,  Gawiser E.,    Prochaska J.~X.,  2005, ARA\&A, 43, 861

\bibitem[\protect\citeauthoryear{{Wolfe}, {Lanzetta}, {Turnshek} \&
  {Oke}}{{Wolfe} et~al.}{1992}]{wolfe92}
{Wolfe} A.~M.,  {Lanzetta} K.~M.,  {Turnshek} D.~A.,    {Oke} J.~B.,  1992,
  ApJ, 385, 151

\bibitem[\protect\citeauthoryear{{Wolfe} \& {Prochaska}}{{Wolfe} \&
  {Prochaska}}{1998}]{wolfe98}
{Wolfe} A.~M.,  {Prochaska} J.~X.,  1998, ApJ, 494, L15

\bibitem[\protect\citeauthoryear{Wolfe, Turnshek, Smith \& Cohen}{Wolfe
  et~al.}{1986}]{wolfe86}
Wolfe A.~M.,  Turnshek D.~A.,  Smith H.~E.,    Cohen R.~D.,  1986, ApJS, 61,
  249

\bibitem[\protect\citeauthoryear{{York}, {Kanekar}, {Ellison} \&
  {Pettini}}{{York} et~al.}{2007}]{york07}
{York} B.~A.,  {Kanekar} N.,  {Ellison} S.~L.,    {Pettini} M.,  2007, MNRAS,
  382, L53

\end{thebibliography}

\label{lastpage}
\end{document}